\documentclass{ws-procs975x65}
\begin{document}
\title{\uppercase{High-$z$ cosmography at a glance}}

\author{\uppercase{Vincenzo Vitagliano}$^{1}$, \uppercase{Jun-Qing Xia}$^2$, \uppercase{Stefano Liberati}$^{3,4}$ \\and \uppercase{Matteo Viel}$^{4,5}$}

\address{$^1$~CENTRA, Departamento de F\'isica, Instituto Superior T\'ecnico,
Universidade T\'ecnica de Lisboa - UTL, Av. Rovisco Pais 1, 1049 Lisboa, Portugal.\\
${}^2$Key Laboratory of Particle Astrophysics, Institute of High Energy Physics,
Chinese Academy of Science, P.O. Box 918-3, Beijing 100049, P.R. China.\\
${}^3$SISSA, Via Bonomea 265, 34136 Trieste, Italy.\\
${}^4$INFN sez. Trieste, Via Valerio 2, 34127 Trieste,
Italy.\\
${}^5$INAF-Osservatorio Astronomico di Trieste, Via
G.B. Tiepolo 11, I-34131 Trieste, Italy.}

\begin{abstract}
Cosmography is the tool that makes possible to untie the 
interpretation of cosmological observations from the definition of any dynamical prior. We review the constraints on the 
cosmographic parameter obtained using the most thorough data set ensemble available. We focus on some specific topics about 
the statistically based selection of the most stringent fitting expansion.
% with all the most powerful cosmological probes:
% Supernovae type Ia and Gamma Ray Bursts, the Baryonic Acoustic Oscillations, the CMB power spectrum 
% and the OHD data. We stress the importance of a statistical method to decide the 
% order of the expansion at which we must stop.
\end{abstract}

\bodymatter

\section{Introduction}

Which kind of information on the cosmological history can be extracted examining 
cosmological observations without the bias of a specific theoretical model? 

Cosmography is the approach to the data analysis that provides the answer to this question.
The assumption on the cosmological metric to fulfill the minimal dynamic request of (spatial) homogeneity and isotropy 
(namely, starting from the working hypothesis that the universe is described by a geometry with Friedmann-Lema\^{\i}tre-Robertson-Walker symmetries)
allows to expand any relevant distance indicator of an observed object in a power series of a suitable redshift parameter.

The coefficients (evaluated today) of such powers can then be expressed in terms of the scale
factor $a(t)$ and its successive derivatives (the cosmographic parameters), supplying the relevant information to describe the kinematic of the universe.

A comment is necessary here: the choice of an appropriate parameter for measuring redshifts is compulsory when data from very distant objects are considered. 
The lack of convergence properties of the series, Taylor-expanded 
in the usual $z$, causes at least two main difficulties: 
an underestimate of the errors and a poor control of the approximation associated with the truncation of the expansion.
These properties can be retrieved recasting all the involved quantities as functions of an
improved parameter $y = z/(1 + z)$ \cite{Visser:2003vq, Cattoen:2007sk}. Being $z \in (0,\infty)$ mapped
into $y\in(0, 1)$, the
convergence of the Taylor series is restored also for high redshifts.

\section{Cosmography beyond standard candles and rulers}

Our recent analysis\cite{Xia:2011iv} handles the problem of interpreting the data
under a cosmographic perspective exploring the whole ensemble of available cosmological data sets. We constrain the parameters appearing in the 
expansions of the characteristic scales associated to these indicators: Supernovae Type Ia (SNeIa) and 
Gamma Ray Bursts (GRBs), Baryon Acoustic Oscillations (BAOs), the Cosmic
Microwave Background (CMB) power spectrum and the direct determination of the Hubble parameter as estimated from surveys of galaxies (Hub).

It is worth noting that, depending on the quantity that has been measured, it could be more appropriate to consider a particular cosmological 
distant scale than another one. To different distant scales correspond different Taylor expansions whose coefficients will combine the cosmographic 
parameters in different ways. For this reason it will be more natural to evaluate some cosmographic parameters in a well defined distance. In our analysis
we refer to the following physical quantities: luminosity distance as the most direct choice for SNeIa and GRBs; 
volume distance for BAOs; angular diameter distance for the CMB. 

The measurement of the Hubble parameter deserves a separate discussion: it can be 
straightforwardly used for cosmographic purposes, but with a caveat. In fact, the coefficient of the $n$-th 
$y$-power in the Taylor expansion already provides a combination of $n$ cosmographic parameters, while the same number of parameters appears only at the
$(n+1)$-th power of the series expansion for the other distant scales. This is essentially due to an extra derivative term appearing in the definition 
of the Hubble parameter. For the different nature of this latter data set, we consider constraints on the cosmographic coefficients initially without the 
Hubble parameter measurements and then adding these data with one order less in the $y$-power expansion.
See Ref. [\refcite{Xia:2011iv}] for further details.

\section{Statistics selection between two truncations: the $F$-test}

It is quite obvious that by adding higher order powers to the redshift expansions it is possible to improve the data fitting, since more free
parameters are involved. However, for a given data set, there will be an upper bound on the order which is statistically significant in
the data analysis\cite{Vitagliano:2009et}. To make matters even worse, we have shown that also an early truncation of the power series
leads to several inconsistencies or artifacts in the analysis of the cosmographic expansion. 
These issues justify the selection of some criteria to properly choose, between two alternatives, the model fitting the data the better and in 
the more statistically significant way.

The solution that can be adopted is the use of a statistical tool, the $F$-test, able to compare two
nested models (in this case, two different truncations of the Taylor
series) in order to find out which is, for a given data set, the most viable approximation
of the series. 
Supposing that the null hypothesis implies the correctness of the first model,
the $F$-test verifies the probability for the
alternative model to fit the data as well. If this probability is
high, then no statistical benefit comes from the extra degrees of
freedom associated to the new model.
The less is this probability, the better is the data fitting of the second model against the
first one.

Quantitatively, the $F$-ratio distribution
among the two polynomials is defined as $F\equiv[(\chi_1^{\,2}-\chi_2^{\,2})/\chi_2^{\,2}]*[(N-n_2)/(n_2-n_1)]$,
where $N$ is the number of data points, and $n_i$ represents the
number of parameters of the $i$-model. Assigned two nested fitting models, and hence specified the $\chi_i^{\,2}$ and the $n_i$, we obtain a specific point
in the $F$-distribution. The integral (also known as $P$-value) of the $F$-distribution curve, lower-bounded by the point of $F$-ratio previously found, 
quantifies the viability of the two matching models. We use the stringent threshold of $5\%$ as the significance level on the $P$-value under which the model 
with one more parameter fits the data better than the other one.

The following Table summarizes the estimates obtained in our analysis for the most statistically meaningful (in the $F$-test 
sense) term of the series expansion.
% 
% The results of our cosmographic analysis are summarized in the Table, showing the estimates obtained for the most statistically meaningful (in the $F$-test 
% sense) term of the expansion.

% \begin{table}\label{table}
% \centering
\begin{center}
\begin{tabular*}{\textwidth}{@{\extracolsep{\fill}}c@{\extracolsep{\fill}}|@{\extracolsep{\fill}}c@{\extracolsep{\fill}}|@{\extracolsep{\fill}}c@{\extracolsep{\fill}}|
@{\extracolsep{\fill}}c@{\extracolsep{\fill}}|@{\extracolsep{\fill}}c@{\extracolsep{\fill}}|@{\extracolsep{\fill}}c@{\extracolsep{\fill}}}
  \hline
  \hline
    Data&\multicolumn{5}{c}{SNIa+GRB+BAO+CMB ($5^{th}$ order)}\\
  \hline
  Parameter&$q_0$&$j_0$&$s_0$&$c_0$&$H_0$\\
  \hline
  Best Fit&$-0.17$&$-6.92$&$-74.18$&$-10.58$&$-$\\
  Mean&$-0.49\pm0.29\,$&$-0.50\pm4.74\,$&$-9.31\pm42.96\,$&$126.67\pm190.15\,$&$-$\\
  \hline
  $\chi^2_{\rm min}/$d.o.f.&\multicolumn{5}{c}{627.61/624}\\
  \hline
  \hline
    Data&\multicolumn{5}{c}{SNIa+GRB+BAO+CMB ($5^{th}$ order) +Hub ($4^{th}$ order)}\\
  \hline
  Parameter&$q_0$&$j_0$&$s_0$&$c_0$&$H_0$\\
  \hline
  Best Fit&$-0.24$&$-4.82$&$-47.87$&$-49.08$&$71.65$\\
  Mean&$-0.30\pm0.16\,$&$-4.62\pm1.74\,$&$-41.05\pm20.90\,$&$-3.50\pm105.37\,$&$71.16\pm3.08\,$\\
  \hline
  $\chi^2_{\rm min}/$d.o.f.&\multicolumn{5}{c}{639.81/633}\\
  \hline
  \hline
\end{tabular*}
\end{center}
% \caption{Constraints on the cosmography parameters up to fifth order
% term from different data combinations.}
% \end{table}

It is interesting to stress that the remarkably good
performance of $\Lambda$CDM, even with respect
to a cosmographic expansion with more free parameters, could be taken as
a strong hint in favor of this specific solution. However, we should warn about the (ab)use of a statistical comparison in terms of the derived $\chi^2$.
In fact, while this procedure is completely meaningful for a selection between two nested cosmographic expansions, it becomes rather questionable when the
comparison is between any fiducial cosmological model (as $\Lambda$CDM) and a cosmographic series, since it
somehow spoils the attempt of tackling the role of standard candles (or rulers) at a more fundamental level as in the spirit of the cosmographic approach.

%Anyway, in the limit in which the series has infinite terms one should converge to the actual cosmology. 

\section*{Acknowledgements}
VV is supported by FCT - Portugal through the grant SFRH/BPD/77678/2011.
JX is supported by the National Youth Thousand Talents Program and
the grant No. Y25155E0U1 from IHEP.
MV is supported by PRIN-MIUR, PRIN-INAF 2009, 
ASI/AAE, INFN/PD-51 grant and the ERC Starting Grant “cosmoIGM”.

\end{document}